\documentstyle[epsf,sprocl]{article}
\newcommand{\beq}    {\begin{equation}}
\newcommand{\eeq}    {\end{equation}}
\newcommand{\beqarr} {\begin{eqnarray}}
\newcommand{\eeqarr} {\end{eqnarray}}
\newcommand{\barr}   {\begin{array}}
\newcommand{\earr}   {\end{array}}

\newcommand{\lsim}{\mathrel{\mathop{\kern 0pt \rlap
  {\raise.2ex\hbox{$<$}}}
  \lower.9ex\hbox{\kern-.190em $\sim$}}}
\newcommand{\gsim}{\mathrel{\mathop{\kern 0pt \rlap
  {\raise.2ex\hbox{$>$}}}
  \lower.9ex\hbox{\kern-.190em $\sim$}}}

\newcommand{\mbi}[1] {\mbox{\scriptsize #1}}

\begin{document}
\title{COSMIC ANTIPROTONS FROM NEUTRALINO ANNIHILATION IN THE GALACTIC HALO}
\author{G. Mignola}
\address{Theoretical Physics Division, CERN, Geneva 23, CH-1211, Switzerland \\
and INFN Sezione di Torino, Via P. Giuria 1, 10125 Turin, Italy}
\maketitle
\begin{abstract}
We investigate the possibility that the antiproton--to--proton flux ratio 
which is measured in cosmic rays
may be generated by neutralino--neutralino annihilation in the galactic halo. 
We study the most general compositions for the relic 
neutralinos. We compare our results with the present experimental sensitivity 
and find that the theoretical predictions are at the level of the
current experimental limits for some regions in the parameter space of the 
model.
We expect that the future measurements of the $\bar p/p$ will provide very 
useful information, complementary to the ones obtainable with other 
experimental means.
\end{abstract}

\section{Introduction}

One of the best motivated particle candidates for cold dark matter
is provided by supersymmetry \cite{EHNOS}.
The lightest supersymmetric particle (LSP), being stable provided R-parity is
conserved, is, in a large region of the supersymmetric parameter space,
the lightest neutralino ($\chi$). This state is defined as the lowest-mass
superposition of the gaugino ($\tilde{B},\tilde{W^3}$) and the higgsino
($\tilde{H_1},\tilde{H_2}$) fields~:
\beq
\chi \; = \; a_1 \tilde{B} + a_2 \tilde{W^3} + a_3 \tilde{H_1} +
a_4 \tilde{H_2} \;\; .
\label{eq:neutr}
\eeq
The possibility of detecting neutralinos in our galactic halo
has been studied extensively in many different ways \cite{GKJ}.
Direct detection of the nuclear recoil induced by neutralino--nucleus
elastic scattering is, both theoretically \cite{th_dir} and experimentally
\cite{exp_dir,this_conf_dir}, under investigation. 
The indirect detection of high-energy
neutrinos coming from the centre of the Earth or the Sun \cite{nu}, or
the indirect detection of exotic components in the primary cosmic rays,
such as $\gamma$, $e^+$ and $\bar{p}$, have been studied in detail
\cite{antip_other,antip_noi}. Here we examine the latter possibility, and
analyse the parameter space of the supersymmetric model taking
care of the new experimental limits on supersymmetric particles coming from
CERN LEP \cite{LEPII}. We also include some recent 
progress on the treatment of the propagation of 
cosmic antiprotons $\bar{p}$ in the Galaxy \cite{CMST}, considering the 
properties of the diffusion model instead of using the standard leaky box 
approach.

\section{Theoretical framework}
We recall in this section only the main features of the model to which we
refer: the Minimal Supersymmetric Standard Model (MSSM).
For a detailed description of the model and for a discussion on
the range of variation of the various parameters that enters in the current 
analysis see: A. Bottino, these Proceedings.

The neutralino mass and the coefficients $a_i$ of Eq.(\ref{eq:neutr})
depend on the parameters: $\mu$ (Higgs mixing parameter), 
$M_1$, $M_2$ (masses of $\tilde{B}$  and of $\tilde{W^3}$, respectively)
and $\tan \beta = v_u/v_d$ ($v_u$ and $v_d$ are 
the v.e.v.'s which give masses to up-type and down-type quarks).
As for the masses of the 
particles entering in the amplitudes for the neutralino-neutralino 
annihilation and for the neutralino-quark scattering 
(Higgs bosons, sfermions), we do 
not make use of theoretical assumptions implied by supergravity schemes, 
but we only take into account constraints due to present experimental 
limits. In the case of the neutral Higgs bosons (the two
CP--even bosons: $h,H$ (of masses $m_h$, $m_H$ with $m_H>m_h$) and the CP--odd
one: $A$ (of mass $m_A$)) we take $m_A$ as a free parameter only bounded by the
present experimental results \cite{higgs}. 
As for the sfermion masses and in order to deal with a relatively small 
number of parameters, 
we consider a common soft scalar mass $m_0$ as a free parameter 
(with the possible exception of the top
squark). Another quantity which we take as a free  parameter is the top 
trilinear coupling $A_t$. 
Moreover we assume the standard GUT relation among the gaugino masses
\beq
M_1:M_2:M_3=\alpha_1:\alpha_2:\alpha_3
\eeq
which implies at $m_Z$ scale that $M_1 \simeq 0.5 M_2$.

Once defined our parameter space, we impose the experimental constraints deduced
from the new particles searches at accelerators, 
including the new lower limits on the
chargino mass obtained at LEPII \cite{LEPII}, and the measurement of the
rare decay $b\rightarrow s\gamma$ \cite{cleo} from CLEO.

We then compute  the neutralino relic abundance in the standard way 
\cite{omega,omega_poles,omega1,ap1}
and we exclude from the parameter space those configurations that do not
fulfil the cosmological bounds $\Omega h^2 \leq 1$.

\subsection{Relic density}
The neutralino relic abundance $\Omega_{\chi} h^2$ is evaluated
following the standard procedure, which gives essentially 
$\Omega_{\chi} h^2 \propto <\sigma_{\mbi{ann}} v>^{-1}_{\mbi{int}}$,
where $<\sigma_{\mbi{ann}} v>_{\mbi{int}}$ is the
thermally--averaged annihilation cross section, integrated from the
freeze--out temperature to the present temperature. The standard
expansion $<\sigma_{\mbi{ann}} v> = a + b x + ...$
may be employed, with $x=T/m_\chi$,
except at s--channel resonances ($Z,A,H,h$) or at the opening of a
new final state channel, where a more precise
treatment has to be used for the thermal average \cite{omega_poles}.
In the evaluation of
$<\sigma_{\mbi{ann}} v>$ the full set of annihilation final
states ($f \bar{f}$ pairs, gauge--boson pairs,
Higgs--boson pairs and Higgs--gauge boson pairs), as well as the
complete set of Born diagrams, are taken into account
\cite{ap1}. 
We note that the constraint $\Omega_{\chi} h^2 \leq 1$ can be very effective
in cutting the parameter space especially for small and intermediate values 
of $\tan \beta$, but is not really restrictive for
large values of $\tan \beta$.

\section{The antiproton spectrum}
Let us turn now to the evaluation of the antiproton signal due to neutralino
annihilation in the galactic halo.
The differential rate (per unit volume and per second) for the production
of antiprotons from $\chi$--$\chi$ annihilations is given by
\beq
S(E_{\bar{p}}) \; = \;
\langle\sigma v\rangle f(E_{\bar p})
\left( \frac{\rho_{\chi}}{m_{\chi}}\right)^2 \;\; ,
\label{susy_source}
\eeq
in which $E_{\bar{p}}$ denotes the antiproton energy, $\sigma$ is the
$\chi$--$\chi$ annihilation cross section and $v$  is the neutralino velocity
in the galactic halo. The neutralino density $\rho_{\chi}$ is a function of the
position $\vec{r}$ in the galactic halo. 
For a single annihilation, the antiproton energy spectrum is
\beq
f(E_{\bar p}) \; \equiv \;
 {1 \over \sigma}  \;
{{d \sigma (\chi \chi \rightarrow \bar{p} + X)} \over {dE_{\bar{p}}}}
 \; = \; {\displaystyle \sum_{F,f}} \;
B^{(F)}_{\chi f} \;\;\;
\left( {dN^{f}_{\bar{p}}\over dE_{\bar{p}}} \right) \;\; ,
\eeq
where $F$ describes the $\chi$--$\chi$ annihilation final state and
$B^{(F)}_{\chi f}$ is the branching ratio into the quarks or gluons $f$ in the
channel $F$. The differential distribution of the antiprotons generated
by the hadronization of quarks (with the exception of the top quark)
and of gluons, is denoted by $dN^{f}_{\bar{p}} / dE_{\bar{p}}$ and
depends on the nature of the species $f$. 
In Eq.(\ref{susy_source}), $\langle\sigma v\rangle$
and $f(E_{\bar{p}})$ depend on the neutralino properties.
The antiproton production rate also depends on the distribution
${\rho}_{\chi}$ of neutralinos inside the galactic
halo. 

For the $\bar{p}$ differential distribution
$f(E_{\bar{p}})$, we have evaluated the branching ratios
$B^{(F)}_{f}$ for all annihilation final states that may produce
antiprotons, i.e. direct production of quarks and gluons,
generation of quarks through the intermediate production of
Higgs bosons, gauge bosons and the top quark. The distributions
$dN^{f}_{\bar{p}}/dE_{\bar{p}}$ from the hadronization
of quarks (with the exception of the top quark) and gluons have been
computed by using the Monte Carlo code JETSET 7.2 \cite{BENGTSSON}.

\noindent The neutralino halo distribution
${\rho}_{\chi}$ is taken to be spherically symmetric and is given by the
standard expression:
\beq
\rho_{\chi} (r) \; = \; \rho_{\chi}(\odot) \;
\frac {a^2 + r_\odot^2}{a^2 + r^2}  \;\; ,
\label{densite_neutralino}
\eeq
where $a = 3$ kpc is the core radius of the dark matter halo.
Particular care must be taken about the local neutralino density
$\rho_{\chi}(\odot)$, which depends on on the
distribution of dark matter in the galactic halo as well as on
the LSP properties. On the one hand the presence of a clumpy
structure over the smooth distribution 
\ref{densite_neutralino} may well enhance the expected signal
\cite{wasserman}.
On the other hand if the big-bang relic density $\Omega_{\chi} h^{2}$,
which we evaluate following the method previously discussed, is too small to
account for the cosmological dark matter, the density of neutralinos
in the galactic halo should be corrected by a factor of
$\xi$. The latter deals with the fact that the neutralino density
is less than the halo density whenever $\Omega_{\chi} h^{2}$
is smaller than a minimal value of, say,
$(\Omega h^2)_{min} = 0.03$, which is compatible with the
observed rotation curve of the Galaxy. At cosmological distances,
this ratio is given by
\beq
\xi \; = \;
{\displaystyle \frac{\Omega_{\chi} h^{2}}{(\Omega h^2)_{min}}} \;\; .
\eeq

\begin{figure}[htb]
\epsfxsize=10cm
\centerline{\epsffile{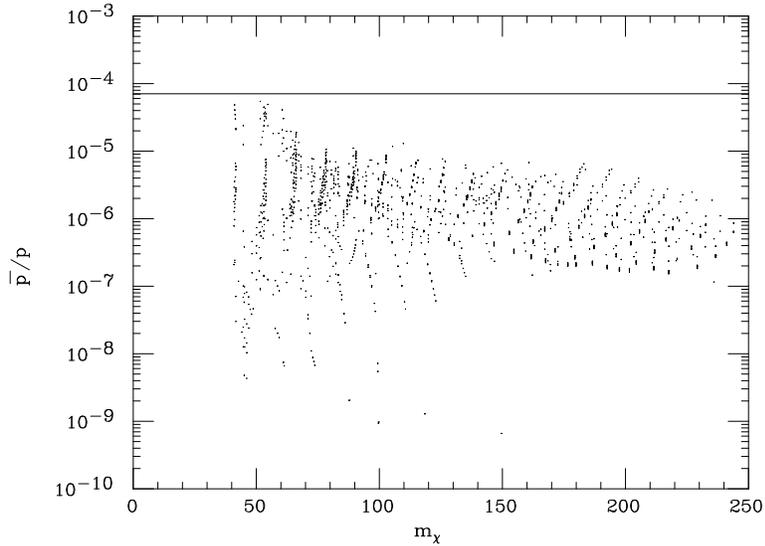}}
\caption{The ratio $\bar{p}/p$ as a function of the neutralino mass for the
simple scan of the parameter space as described in the text. The horizontal
line represent the current experimental upper bound \protect\cite{IMAX}.}
\end{figure}

The antiproton flux expected at Earth can be written as

\beq
 \Phi_{\bar p}(E_{\bar p}) =
{ 1 \over { 4 \pi}} <\sigma v> f(E_{\bar p}) 
\left ( {{\bar {\rho}_{\chi}} \over {m_\chi}}
\right )^2 v_{\bar p} \tau _{\bar p}(E_{\bar p}) 
\eeq

\noindent where $v_{\bar{p}}$ and  $\tau_{\bar{p}}(E_{\bar{p}})$ are the $\bar{p}$ 
velocity and confinement time. 

As far as $\tau_{\bar{p}}(E_{\bar{p}})$ is concerned we employ the
results of Ref.\cite{CMST} and we introduce an energy dependence of
$\tau_{\bar{p}}$ as deduced from a detailed calculation a diffusion model
to treat the $\bar{p}$ propagation.

\begin{figure}[htb]
\epsfxsize=10cm
\centerline{\epsffile{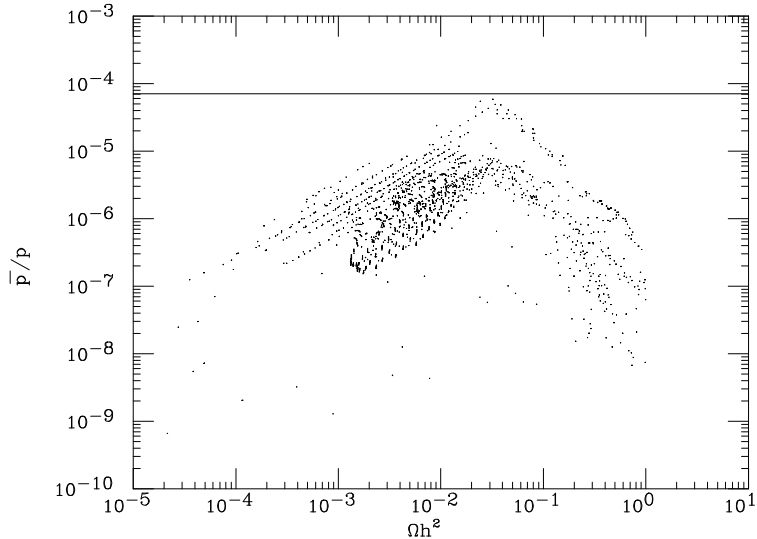}}
\caption{The ratio $\bar{p}/p$ as a function of the neutralino relic density
$\Omega_\chi h^2$ for the
simple scan of the parameter space as described in the text. The horizontal
line represent the current experimental upper bound \protect\cite{IMAX}.}
\end{figure}

To compare predictions with experimental data we have to take into account the
solar modulation of the antiprotons as well as the modulation of the primary
protons. We have used the results of Ref. \cite{perko} changing the values of 
the parameters of the modulation model in order to properly compare with the
experiments \cite{IMAX}. 

\section{Results} 
In Fig.1 we display the ratio $\bar{p}/p$ as a function of the neutralino
mass for a simple scan of the parameter space, where  $\tan \beta=1.1$ and 
$\tan \beta=55$, and $m_A = 500$ GeV and $m_A = 50 - 60$ GeV.
This choice
represents almost the two boundaries for the range of variation of these
parameters.   The horizontal
line represents the present level of sensitivity in the energy range 
$0.25~\mbox{GeV} \leq T \leq 1$ GeV as measured by the IMAX 
collaboration \cite{IMAX}, 
that is
\beq
\left(\frac{\bar{p}}{p} \right) \leq 7.1 \times 10^{-5}~\mbox{95 \% C.L.}~.
\eeq
We notice that for some configurations of the model we are at the
level of the experimental result.

\begin{figure}[htb]
\epsfxsize=10cm
\centerline{\epsffile{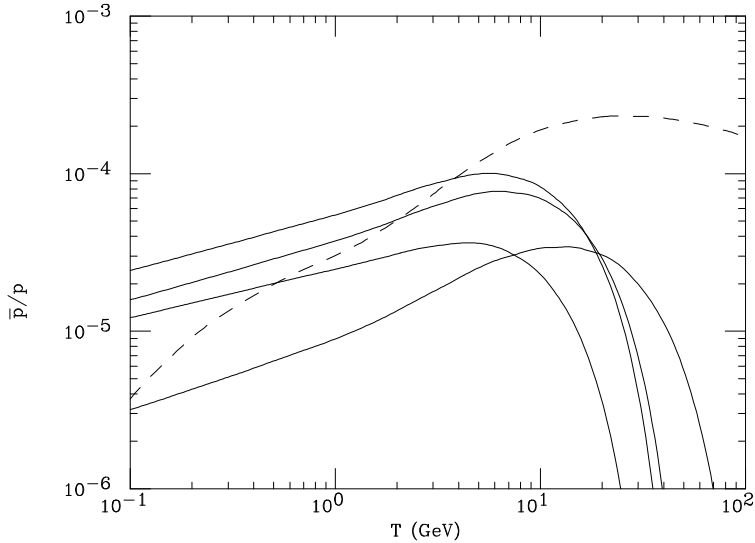}}
\caption{The ratio $\bar{p}/p$ as a function of the kinetic energy $T$.  
The solid line represent four different values of the $\chi$ mass (from left to
right: $40$ GeV, $50$ GeV, $60$ GeV, $120$ GeV). The dashed line represent the
expected background due to secondary production of $\bar{p}$ 
\protect \cite{CMST}.}
\end{figure}

In Fig.2 we show the $\bar{p}/p$ flux versus the cosmological relic
density $\Omega_\chi h^2$. This plot illustrates the dependence of the signal on
the rescaling procedure we have adopted, here we recall that the value $(\Omega
h^2)_{min} = 0.03$ has been used.

In Fig.3 we display the $\bar{p}/p$ ratio as a function of the
kinetic energy $T$. The different curves represent four different neutralino
masses and compositions. In the same plot is shown, as a dotted line, 
the expected $\bar{p}/p$ background of the secondary production by spallation 
of the primary cosmic ray flux (taken from Ref. \cite{CMST}). 

A possible clear signature to discriminate  signal over background is the energy
dependence of the two fluxes on $T$: the former being a milder function of T at
low energies with respect to the latter, as it is clearly seen in Fig. 3. This
methods will be useful at the forthcoming generation of experiments such as
AMS \cite{AMS} or PAMELA \cite{pam} that will collect a much larger 
number of $\bar{p}$ with respect to the present experimental situation.

In summary we have shown that the production of $\bar{p}$'s from
$\chi$-$\chi$ annihilation in our galactic halo is, in some regions of
the MSSM parameter space, at the level of the present experimental sensitivity.
The forthcoming generation of detectors will be able in the near future to
investigate large regions of the MSSM by discriminating the signal over the
background with an analysis of the energy dependence of the 
measured $\bar{p}/p$.

\noindent{\bf Acknowledgements}.
I wish to thank Sandro Bottino, Fiorenza Donato, Nicolao Fornengo and 
Stefano Scopel for the invaluable help and advice in doing this work. 
The present investigation was performed as a part of a wider project to be
carried out in a collaboration among researchers of the Stockholm University,
the LAPP (Annecy) and the Torino University. I wish also to acknowledge a
fellowship of the Torino University .

\end{document}